\documentclass[12pt]{article}

\usepackage{latexsym}

\usepackage{graphicx}

\begin{document}


\title{Solution generating in 5D Einstein-Maxwell-dilaton
 gravity and derivation of dipole black ring solutions}

\author{
     Stoytcho S. Yazadjiev \thanks{E-mail: yazad@phys.uni-sofia.bg}\\
{\footnotesize  Department of Theoretical Physics,
                Faculty of Physics, Sofia University,}\\
{\footnotesize  5 James Bourchier Boulevard, Sofia~1164, Bulgaria }\\
}

\date{}

\maketitle

\begin{abstract}
We consider 5D Einstein-Maxwell-dilaton  (EMd) gravity in spacetimes with
three commuting Killing vectors: one timelike and two spacelike Killing vectors, one
of which is  hypersurface-orthogonal. Assuming a special ansatz for the Maxwell field
we show that the 2-dimensional reduced EMd equations are completely integrable. We also develop a solution
generating method for explicit construction of exact EMd
solutions from known exact solutions of 5D vacuum Einstein equations with considered symmetries.
We derive explicitly the rotating dipole black ring solutions as a particular application of the
solution generating method.
\end{abstract}


\sloppy

\section{Introduction}

In recent years the higher dimensional gravity is attracting much interest. Apart from the fact
that the higher dimensional gravity is interesting in its own right, the increasing amount of works
devoted to the study of the higher dimensional spacetimes is inspired from the string theory and
the brane-world scenario with large extra dimensions.
The gravity in higher dimensions exhibits much richer dynamics and spectrum  of
solutions than in four dimensions. One of the most reliable routes for  better understanding of higher
dimensional gravity and the related topics are the exact solutions. However, the higher dimensional solutions
found so far are not so many. As yet to the best of our knowledge
there are no EMd solutions found in the literature that describe rotating charged black holes in
higher dimensions with an arbitrary dilaton coupling parameter (there are some numerical solutions, however \cite{Kunz1,Kunz2}).
Moreover, unlike the 4D case, in higher dimensions  the systematic construction
of new solutions has not been accomplished yet.
It is well known that both vacuum and electrovacuum 4D Einstein equations  are completely integrable
being restricted to spacetimes with two-dimensional Abelian group of isometries \cite{RG}-\cite{N}.
This nice property is also shared  by some effective string gravity models (or certain sectors of them) which allows us
to find many families of physically interesting exact solutions \cite{B}-\cite{YU}.
The $D$-dimensional vacuum Einstein equations with $(D-2)$-dimensional Abelian group of
isometries are completely integrable, too  \cite{DM2},\cite{POM}. Recently, in \cite{Y},
we have shown that, after imposing some symmetries on the spacetime and the electromagnetic filed,
the 5D Einstein-Maxwell equations are completely integrable.

The aim of this work is to make a step towards the systematic construction of exact solutions in 5D
Einstein-Maxwell-dilaton (EMd) gravity. We consider 5D EMd gravity in spacetimes with
three commuting Killing vectors: one timelike and two spacelike Killing vectors, one of
which is hypersurface-orthogonal. Assuming a special ansatz for the Maxwell field
we show that the 2-dimensional reduced EMd equations are completely integrable by deriving a
Lax-pair presentation. We also develop a solution generating method for explicit construction of exact EMd
solutions with considered symmetries.
The rotating dipole black ring solutions in EMd gravity are derived as a particular application of the
developed solution generating method as well.

\section{Dimensional reduction, coset presentation and  complete integrability }

We consider the 5D EMd gravity described by the action

\begin{eqnarray}\label{A}
S= {1\over 16\pi} \int d^{5}x \sqrt{-g}\left[R - 2g^{\mu\nu}\partial_{\mu}\varphi\partial_{\nu}\varphi
- {1\over 4} e^{-2\alpha\varphi} F_{\mu\nu}F^{\mu\nu} \right]
\end{eqnarray}

where $\alpha\ne 0$ is the dilaton coupling parameter.

This action  yields the following field equations

\begin{eqnarray}\label{EMdFE}
R_{\mu\nu} &=& 2\partial_{\mu}\varphi \partial_{\nu}\varphi
+ {1\over 2}e^{-2\alpha\varphi} \left(F_{\mu\lambda}F_{\nu}^{\,\lambda} - {1\over 6} F_{\sigma\lambda}F^{\sigma\lambda} g_{\mu\nu}\right) ,
 \nonumber \\
\nabla_{\mu}\nabla^{\mu}\varphi &=& - {\alpha \over 8} e^{-2\alpha\varphi} F_{\sigma\lambda}F^{\sigma\lambda} , \\
&&\nabla_{\mu}\left(e^{-2\alpha\varphi} F^{\mu\nu} \right) = 0 , \nonumber
\end{eqnarray}

In this paper we consider  spacetimes with  three commuting Killing vectors:
one timelike Killing vector $T$ and two spacelike Killing vectors $K_{1}$ and $K_{2}$.  We also assume
that the Killing vector $K_{2}$ is hypersurface orthogonal. We require the electromagnetic and dilaton fields  to be
invariant under the Abelian group generated by the Killing vectors, i.e.

\begin{eqnarray}
L_{K_{1}}F = L_{K_{2}}F=L_{T}F = 0 ,\,\,\,\,\,\,  L_{K_{1}}\varphi = L_{K_{2}}\varphi=L_{T}\varphi = 0 ,
\end{eqnarray}

where $L_{K}$ denotes the Lie derivative along the vector $K$.

In adapted coordinates in which $K_{2}=\partial/\partial Y$, the spacetime
metric can be written in the form

\begin{equation}
ds^2 = e^{2u}dY^2 + e^{-u} h_{ij}dx^idx^j
\end{equation}

where $h_{ij}$ is a $4$-dimensional metric with Lorentz signature. Both $u$ and $h_{ij}$
depend on the coordinates $x^i$ only. The electromagnetic field is taken in the form

\begin{equation}
F = dA_{Y}\wedge dY
\end{equation}

where $A_{Y}$ depends on the coordinates $x^{i}$ only. Let us note that this form of the electromagnetic field is compatible
with the invariance of $F$ with respect to the Killing vectors.

After a dimensional reduction along the Killing vector $K_{2}$, the field equations (\ref{EMdFE})
are reduced to the following effective 4D theory:

\begin{eqnarray}
&&{\cal D}_{i}{\cal D}^{i}u =
- {1\over 3} e^{-2u-2\alpha\varphi}h^{ij}{\cal D}_{i}A_{Y} {\cal D}_{j}A_{Y},\\
&&{\cal D}_{i}{\cal D}^{i}\varphi = - {\alpha\over 4} e^{-2u-2\alpha\varphi}h^{ij}{\cal D}_{i}A_{Y} {\cal D}_{j}A_{Y} ,\\
&&{\cal D}_{i}\left(e^{- 2u-2\alpha\varphi}{\cal D}^{i}A_{Y} \right) = 0, \\
&&R_{ij}(h)= {3\over 2}\partial_{i}u\partial_{j}u + 2\partial_{i}\varphi\partial_{j}\varphi
+ {1\over 2}e^{-2u-2\alpha\varphi}\partial_{i}A_{Y}\partial_{j}A_{Y}.
\end{eqnarray}

Here ${\cal D}_{i}$ and $R_{ij}(h)$ are the covariant derivative and Ricci tensor with respect
to the  Lorentz metric $h_{ij}$. We shall introduce the new parameter $\alpha_{*}$ and the fields $\chi$ and $\zeta$,
defined  by

\begin{eqnarray}
\alpha_{*} &=& {\sqrt{3}\over 2}\alpha ,\\
\chi &=& u + \alpha\varphi ,\\
\zeta &=& u - {2\over \sqrt{3}\alpha_{*} }\varphi.
\end{eqnarray}

Further we  introduce the symmetric matrix $M_{1}$ given by

\begin{eqnarray}
M_{1} = \left(%
\begin{array}{cc}
  e^{\chi} + {1+ \alpha^2_{*}\over 3}e^{-\chi}A^2_{Y} & \sqrt{{1+ \alpha^2_{*}\over 3}} e^{-\chi}A_{Y} \\
 \sqrt{{1+ \alpha^2_{*}\over 3}} e^{-\chi}A_{Y} & e^{-\chi} \\\end{array}%
\right)
\end{eqnarray}

with $\det M_{1}=1$. Then the dimensionally reduced EMd equations become

\begin{eqnarray}
&&{\cal D}_{i}\left[{\cal D}^{i}M_{1}M^{-1}_{1}\right]=0 ,\\
&& {\cal D}_{i}{\cal D}^{i}\zeta = 0,\\
&&R_{ij}(h) =  -{3\over 4(1+ \alpha^2_{*})} Tr\left[\partial_{i}M_{1}\partial_{j}M_{1}^{-1}\right] +
{3\alpha^2_{*}\over 2(1+ \alpha^2_{*}) }\partial_{i}\zeta \partial_{j}\zeta .
\end{eqnarray}

These equations are yielded by the action

\begin{eqnarray}
S = {1\over 16\pi} \int d^4x \sqrt{-h} \left[R(h) + {3\over 4(1+ \alpha^2_{*})}h^{ij} Tr\left(\partial_{i}M_{1}\partial_{j}M_{1}^{-1}\right)
- {3\alpha^2_{*}\over 2(1+ \alpha^2_{*}) }h^{ij}\partial_{i}\zeta \partial_{j}\zeta \right].
\end{eqnarray}

Clearly the action is invariant under the $SL(2,R)\times R$ group where the group action is given by

\begin{eqnarray}
M_{1} \to GM_{1}G^{T}, \,\,\, \zeta \to \zeta + constant ,
\end{eqnarray}

$G \in SL(2,R)$. In fact the matrices $M_{1}$ parameterize a $SL(2,R)/SO(2)$ coset. So we obtain a non-linear
$\sigma$-model coupled to 4D Einstein  gravity with a minimally coupled scalar field $\zeta$.

Next step is to further reduce the effective $4D$ theory along the Killing vectors $T$ and $K_{1}$.
For this purpose, it is useful to introduce the twist of the Killing vector $T$

\begin{eqnarray}\label{TWIST}
\omega = - {1\over 2} \star(h)\left(T\wedge dT \right)
\end{eqnarray}

were $\star(h)$ is the Hodge dual with respect to the metric $h_{ij}$.

One can show that the Ricci 1-form ${\Re}_{h}[T]$  defined by
\begin{equation}
{\Re}_{h}[T] = R_{ij}(h)T^{j}dx^{i} ,
\end{equation}

 satisfies

\begin{equation}
\star(h)\left( T\wedge {\Re}_{h}[T] \right) = d\omega .
\end{equation}

In our case we obatin

\begin{eqnarray}
{\Re}_{h}[T] = {3\over 2}\left(T^{j}\partial_{j}u\right) du + \left(T^{j}\partial_{j}\varphi\right) d\varphi +
{1\over 2} e^{-2u-2\alpha\varphi}\left(T^{j}\partial_{j}A_{Y}\right) dA_{Y}.
\end{eqnarray}

Since the potentials $u$, $\varphi$ and $A_{Y}$ are invariant under the Killing symmetry generated by $T$
we have

\begin{eqnarray}
T^{j}\partial_{j}u = T^{j}\partial_{j}\varphi = T^{j}\partial_{j}A_{Y} = 0
\end{eqnarray}

which gives  ${\Re}_{h}[T]=0$, i.e. $d\omega$=0. Therefore there exists (locally) a  potential $f$ such that

\begin{equation}\label{EFORM}
\omega = df.
\end{equation}

In adapted coordinates for the Killing vectors $T=\partial/\partial t$ and $K_{1}=\partial/\partial X$,
and in the canonical coordinates $\rho$ and $z$ for the transverse space, the 4D metric $h_{ij}$ can be written in the form

\begin{eqnarray}
h_{ij}dx^idx^j = -e^{2U}\left(dt + {\cal A} dX \right)^2 + e^{-2U}\rho^2 dX^2 + e^{-2U}e^{2\Gamma}(d\rho^2 + dz^2).
\end{eqnarray}

For this form of the metric $h_{ij}$, combining (\ref{TWIST}) and (\ref{EFORM}), and  after some algebra we find
that  the twist potential $f$ satisfies

\begin{eqnarray}\label{TPS}
\partial_{\rho}f &=& -{1\over 2} {e^{4U}\over \rho} \partial_{z}{\cal A} ,\\
\partial_{z} f &=& {1\over 2} {e^{4U}\over \rho} \partial_{\rho}{\cal A}.
\end{eqnarray}

Before writing the 2D reduced equations  we shall introduce the symmetric matrix

\begin{eqnarray}
M_{2} = \left(%
\begin{array}{cc}
  e^{2U} + 4f^2e^{-2U} & 2fe^{-2U} \\
 2fe^{-2U} & e^{-2U} \\\end{array}%
\right)
\end{eqnarray}

with $\det M_{2}=1$. Then the 2D reduced EMd equations read

\begin{eqnarray}
&&\partial_{\rho}\left(\rho\partial_{\rho}M_{1} M^{-1}_{1} \right)
 + \partial_{z}\left(\rho\partial_{z}M_{1} M^{-1}_{1} \right) = 0 ,\\
&&\partial_{\rho}\left(\rho \partial_{\rho}M_{2}M^{-1}_{2} \right)
 + \partial_{z}\left(\rho \partial_{z}M_{2}M^{-1}_{2} \right) = 0 ,\\
&& \partial_{\rho}\left( \rho\partial_{\rho}\zeta \right)
+ \partial_{z}\left(\rho\partial_{z}\zeta \right) =0 ,\\
\rho^{-1} \partial_{\rho}\Gamma &=&
- {1\over 8} \left[Tr\left(\partial_{\rho}M_{2}\partial_{\rho}M^{-1}_{2}\right)
 - Tr\left(\partial_{z}M_{2}\partial_{z}M^{-1}_{2}\right) \right]  \nonumber \\
&&- {3\over 8(1+ \alpha^2_{*})} \left[Tr\left(\partial_{\rho}M_{1}\partial_{\rho}M^{-1}_{1}\right)
- Tr\left(\partial_{z}M_{1}\partial_{z}M^{-1}_{1}\right) \right] \nonumber \\
&& + {3\alpha^2_{*}\over 4(1+\alpha^2_{*}) }\left[(\partial_{\rho}\zeta)^2 - (\partial_{z}\zeta)^2  \right],\\
\rho^{-1} \partial_{z}\Gamma &=& - {1\over 4} Tr\left(\partial_{\rho}M_{2}\partial_{z}M^{-1}_{2}\right)
\nonumber \\
&& - {3\over 4(1+ \alpha^2_{*})} Tr\left(\partial_{\rho}M_{1}\partial_{z}M^{-1}_{1}\right)
+ {3\alpha^2_{*}\over 2(1+\alpha^2_{*}) } \partial_{\rho}\zeta\partial_{z}\zeta .
\end{eqnarray}

As a result we find that the "field variables" $M_{1}$ and $M_{2}$ satisfy the equations of two
$SL(2,R)/SO(2)$ $\sigma$-models in two dimensions, modified by the presence of the factor $\rho$.
The system equations for $\Gamma$
can be integrated, once a pair of solutions for the two $\sigma$-models  and a solution of the linear equation
for $\zeta$ are known. Therefore,
the problem of generating solutions to the 5D EMd equations with the described symmetries reduces to
the solutions of the two $\sigma$-models and the choice of a harmonic function.

It is well known  that the $\sigma$-model equations are completely integrable \cite{BZ1,BZ2}. This is a consequence
of the fact that the $\sigma$-model equations can be considered as the compatibility condition of
the linear differential equations (Lax-pair presentation)\cite{BZ1,BZ2}

\begin{eqnarray}\label{LPP}
 D_{\rho} \Psi &=& {\rho {\cal U}+ \lambda V\over \lambda^2 + \rho^2} \Psi ,\\
 D_{z} \Psi &=& {\rho V - \lambda {\cal U}\over \lambda^2 + \rho^2} \Psi ,
\end{eqnarray}

where

\begin{eqnarray}
D_{\rho} = \partial_{\rho}  + {2\lambda\rho \over \lambda^2 + \rho^2}\partial_{\lambda}, \,\,\,\,
D_{z} = \partial_{z}  - {2\lambda^2 \over \lambda^2 + \rho^2}\partial_{\lambda}.
\end{eqnarray}

Here $V=\rho\partial_{z}M M^{-1}$, ${\cal U}=\rho\partial_{\rho}M M^{-1}$ and $\lambda$ is the complex
spectral parameter. The "wave function"  $\Psi(\rho,z,\lambda)$ is
a complex matrix. The $\sigma$-model equations then follows from the compatibility condition

\begin{equation}
[D_{\rho}, D_{z}]\Psi = 0.
\end{equation}

The matrix $M$ can be found from the "wave function" $\Psi$ as $M(\rho,z)=\Psi(\rho,z,\lambda=0)$.

The inverse scattering transform (IST) method can be directly applied to (\ref{LPP}) to generate multisoliton
solutions. The dressing procedure allows us to generate new solutions from known ones. Since this dressing
technique is well known we will not discuss it here and refer the reader to \cite{BZ1,BZ2}.

In this paper we will not apply the IST method. In the next section we  present new and simple enough
solution generating method which allows us to generate new 5D EMd solutions from known solutions of the
5D vacuum  Einstein equations.

\section{Solution construction}

Let us consider two solutions $M_{1}=M^{(1)}$ and $M_{2}=M^{(2)}$ of the $\sigma$-model equation

\begin{equation}
\partial_{\rho}\left(\rho\partial_{\rho}M M^{-1} \right)
+ \partial_{z}\left(\rho\partial_{z}M M^{-1} \right) = 0 .\\
\end{equation}

In addition let us denote by $\gamma^{(i)}$ the solution of the system

\begin{eqnarray}
\rho^{-1} \partial_{z}\gamma^{(i)} &=&
-{1\over 4} Tr\left(\partial_{\rho}M^{(i)}\partial_{z}{M^{(i)}}^{-1} \right), \\
\rho^{-1} \partial_{\rho}\gamma^{(i)} &=&
-{1\over 8} \left[Tr\left(\partial_{\rho}M^{(i)}\partial_{\rho}{M^{(i)}}^{-1} \right)
- Tr\left(\partial_{z}M^{(i)}\partial_{z}{M^{(i)}}^{-1} \right) \right].
\end{eqnarray}

Then we find for the metric function $\Gamma$

\begin{equation}
\Gamma = \gamma^{(2)} + {3\over 1 + \alpha^2_{*}}\gamma^{(1)} + {3\alpha^2_{*}\over 1 + \alpha^2_{*}} \nu_{\zeta}
\end{equation}

where $\nu_{\zeta}$ is a solution of the system

\begin{eqnarray}
\rho^{-1}\partial_{\rho}\nu_{\zeta} &=& {1\over 4}\left[\left(\partial_{\rho} \zeta\right)^2
- \left(\partial_{z} \zeta \right)^2 \right],\\
\rho^{-1}\partial_{z} \nu_{\zeta}&=& {1\over 2} \partial_{\rho} \zeta\partial_{z} \zeta.
\end{eqnarray}

From a practical point of view it is more convenient  to associate the $\sigma$-model solutions
$M^{(i)}$  with the vacuum Einstein solutions\footnote{From now on all quantities
with subscript or superscript "E" correspond to the vacuum case.}

\begin{eqnarray}
ds^2_{E(i)} = e^{2u^{(i)}_{E}}dY^2 + e^{-u^{(i)}_{E}}\left [-e^{2{U^{(i)}_{E}}}\left(dt + {\cal A}^{(i)}_{E} dX \right)^2
\right. \\ \left. +  e^{-2{U^{(i)}_{E}}}\rho^2 dX^2 + e^{-2{U^{(i)}_{E}}}e^{2\Gamma^{(i)}_{E}}(d\rho^2 + dz^2)\right]\nonumber .
\end{eqnarray}

which correspond to the matrixes

\begin{eqnarray}
M^{(i)} = \left(%
\begin{array}{cc}
  e^{2U^{(i)}_{E}} + 4\left(f^{(i)}_{E}\right)^2e^{-2U^{(i)}_{E}} & 2f^{(i)}_{E}e^{-2U^{(i)}_{E}} \\
 2f^{(i)}_{E}e^{-2U^{(i)}_{E}} & e^{-2U^{(i)}_{E}} \\\end{array}%
\right) .
\end{eqnarray}

Here the metric function $\Gamma^{(i)}_{E}$ for the vacuum Einstein equations
satisfies\footnote{Obviously, these equations are obtained from the EMd equations by setting $A_{Y}=0$ and $\varphi=0$. }

\begin{eqnarray}
\rho^{-1}\partial_{\rho}\Gamma^{(i)}_{E} &=& -{1\over 8} \left[Tr\left(\partial_{\rho}M^{(i)}\partial_{\rho}{M^{(i)}}^{-1} \right)
- Tr\left(\partial_{z}M^{(i)}\partial_{z}{M^{(i)}}^{-1} \right) \right] \nonumber \\
&& + {3\over 4}\left[\left(\partial_{\rho}u^{(i)}_{E}\right)^2 -
\left(\partial_{z}u^{(i)}_{E}\right)^2 \right] ,\\
\rho^{-1}\partial_{z}\Gamma^{(i)}_{E} &=& -{1\over 4} Tr\left(\partial_{\rho}M^{(i)}\partial_{z}{M^{(i)}}^{-1} \right)
+ {3\over 2} \partial_{\rho} u^{(i)}_{E}\partial_{z}u^{(i)}_{E}.
\end{eqnarray}

It is not difficult to see from these equations that the metric function $\Gamma^{(i)}_{E}$ can be decomposed
into the form

\begin{equation}
\Gamma^{(i)}_{E} = \gamma^{(i)} + \Omega^{(i)}_{E}
\end{equation}

where $\Omega^{(i)}_{E}$ is a solution to the system

\begin{eqnarray}\label{OS}
\rho^{-1}\partial_{\rho}\Omega^{(i)}_{E} &=& {3\over 4}\left[\left(\partial_{\rho} u^{(i)}_{E}\right)^2
- \left(\partial_{z} u^{(i)}_{E}\right)^2 \right],\\
\rho^{-1}\partial_{z}\Omega^{(i)}_{E} &=& {3\over 2} \partial_{\rho} u^{(i)}_{E}\partial_{z} u^{(i)}_{E}.
\end{eqnarray}

Then we find

\begin{equation}
\Gamma = \Gamma^{(2)}_{E} - \Omega^{(2)}_{E} + {3\over 1 + \alpha^2_{*}}\left[\Gamma^{(1)}_{E} - \Omega^{(1)}_{E} + \alpha^2_{*} \nu_{\zeta}\right].
\end{equation}

Further, comparing the matrixes $M_{1}$ and $M^{(1)}$ we obtain

\begin{eqnarray}\label{UAD}
e^{2u} &=& e^{4U^{(1)}_{E}\over 1+ \alpha^2_{*} } e^{{2\alpha^2_{*} \over 1 + \alpha^2_{*} }\zeta} \nonumber,\\
e^{2\alpha\varphi} &=&  e^{{4\alpha^2_{*}\over 1+ \alpha^2_{*} }U^{(1)}_{E}} e^{-{2\alpha^2_{*} \over 1 + \alpha^2_{*} }\zeta}  ,\\
A_{Y} &=& {2\sqrt{3}\over \sqrt{1+ \alpha^2_{*} }} f^{(1)}_{E} \nonumber.
\end{eqnarray}

Having once the metric functions, we  find for the metric

\begin{eqnarray}
ds^2 =  e^{4U^{(1)}_{E}\over 1+ \alpha^2_{*} } e^{{2\alpha^2_{*} \over 1 + \alpha^2_{*} }\zeta} dY^2  +
e^{-{2U^{(1)}_{E}\over 1+ \alpha^2_{*} }} e^{-{\alpha^2_{*} \over 1 + \alpha^2_{*} }\zeta}
\left[- e^{2U^{(2)}_{E}}\left(dt + {\cal A}^{(2)}_{E} dX \right)^2 + e^{-2U^{(2)}_{E}}\rho^2 dX^2 \right. \nonumber \\
\left. + \left({e^{2\Gamma^{(1)}_{E}} \over e^{2\Omega^{(1)}_{E} + {2\over 3}\Omega^{(2)}_{E}} } \right)^{3\over 1+ \alpha^2_{*}}
\left({ e^{2\nu_{\zeta}}\over e^{{2\over 3}\Omega^{(2)}_{E} }} \right)^{3\alpha_{*}^2\over 1+ \alpha^2_{*}}
e^{2\Gamma^{(2)}_{E}}(d\rho^2 + dz^2) \right].
\end{eqnarray}

Taking into account that

\begin{eqnarray}
g^{E(i)}_{00} &=& - e^{-u^{(i)}_{E}}e^{2U^{(i)}_{E}},\\
{\tilde g}^{E(i)}_{XX} &=& g^{E(i)}_{XX} - g^{E(i)}_{00}({\cal A}^{(i)}_{E})^2 = e^{-u^{(i)}_{E}}e^{-2U^{(i)}_{E}}\rho^2,\\
g^{E(i)}_{\rho\rho} &=& e^{-u^{(i)}_{E}}e^{-2U^{(i)}_{E}} e^{2\Gamma^{(i)}_{E}}  ,
\end{eqnarray}

and

\begin{eqnarray}
e^{4U^{(i)}_{E}} &=& (g^{E(i)}_{00})^2 g^{E(i)}_{YY},\\
e^{2\Gamma^{(i)}_{E}} &=& |g^{E(i)}_{00}|g^{E(i)}_{YY} g^{E(i)}_{\rho\rho},
\end{eqnarray}

the metric can be presented in the form

\begin{eqnarray}
ds^2 &=& \left[|g^{E(1)}_{00}|\sqrt{g^{E(1)}_{YY}} \right]^{2\over 1+ \alpha^2_{*}}  e^{{2\alpha^2_{*}\over 1+ \alpha^2_{*}}\zeta}dY^2
 \nonumber \\ &&+ {\sqrt{g^{E(2)}_{YY}} e^{-{\alpha^2_{*}\over 1
+ \alpha^2_{*}}\zeta}\over \left[|g^{E(1)}_{00}|\sqrt{g^{E(1)}_{YY}}\right]^{1\over 1+ \alpha^2_{*}} }
\left[g^{E(2)}_{00}\left(dt + {\cal A}^{(2)}_{E}dX \right)^{2} +
{\tilde g}^{E(2)}_{XX}dX^2  \nonumber \right. \\ && \left. +
\left(|g^{E(1)}_{00}|g^{E(1)}_{YY} g^{E(1)}_{\rho\rho}
\over e^{2\Omega^{(1)}_{E}
+ {2\over 3}\Omega^{(2)}_{E}} \right)^{3\over 1+ \alpha^2_{*}}
\left({e^{2\nu_{\zeta}} \over e^{{2\over 3}\Omega^{(2)}_{E}}}\right)^{3\alpha^2_{*}\over 1
+ \alpha^2_{*}} g^{E(2)}_{\rho\rho} (d\rho^2 + dz^2) \right] .
\end{eqnarray}

Summarizing, we obtain the following  important result presented as a proposition.

{\bf Proposition.} {\it Let us consider two solutions of the vacuum
5D Einstein equations}

\begin{eqnarray}
ds_{E(i)}^2 = g^{E(i)}_{YY} dY^2 + g^{E(i)}_{00}\left(dt + {\cal A}^{(i)}_{E}dX \right)^{2} +
{\tilde g}^{E(i)}_{XX}dX^2 + g^{E(i)}_{\rho\rho} (d\rho^2 + dz^2)
\end{eqnarray}

{\it and a harmonic function $\zeta$}.

{\it Then the following give a solution to the 5D EMd equations\footnote{More generally
we can take $A_{Y}= \pm {2\sqrt{3}\over \sqrt{1 + \alpha^2_{*}}} f^{(1)}_{E}  + const $.} }

\begin{eqnarray}
ds^2 &=& \left[|g^{E(1)}_{00}|\sqrt{g^{E(1)}_{YY}} \right]^{2\over 1+ \alpha^2_{*}}  e^{{2\alpha^2_{*}\over 1+ \alpha^2_{*}}\zeta}dY^2
 \nonumber \\ &&+ {\sqrt{g^{E(2)}_{YY}} e^{-{\alpha^2_{*}\over 1
+ \alpha^2_{*}}\zeta}\over \left[|g^{E(1)}_{00}|\sqrt{g^{E(1)}_{YY}}\right]^{1\over 1+ \alpha^2_{*}} }
\left[g^{E(2)}_{00}\left(dt + {\cal A}^{(2)}_{E}dX \right)^{2} +
{\tilde g}^{E(2)}_{XX}dX^2  \nonumber \right. \\ && \left. +
\left(|g^{E(1)}_{00}|g^{E(1)}_{YY} g^{E(1)}_{\rho\rho}
\over e^{2\Omega^{(1)}_{E}
+ {2\over 3}\Omega^{(2)}_{E}} \right)^{3\over 1+ \alpha^2_{*}}
\left({e^{2\nu_{\zeta}} \over e^{{2\over 3}\Omega^{(2)}_{E}}}\right)^{3\alpha^2_{*}\over 1
+ \alpha^2_{*}} g^{E(2)}_{\rho\rho} (d\rho^2 + dz^2) \right] ,\\
A_{Y} &=& {2\sqrt{3}\over \sqrt{1 + \alpha^2_{*}}} f^{(1)}_{E}, \\
e^{2\alpha\varphi} &=& \left[|g^{E(1)}_{00}|\sqrt{g^{E(1)}_{YY}} \right]^{2\alpha^2_{*}\over 1+ \alpha^2_{*}}  e^{-{2\alpha^2_{*}\over 1+ \alpha^2_{*}}\zeta} ,
\end{eqnarray}

{\it where  $f^{(1)}_{E}$ is a solution  to the  system  }

\begin{eqnarray}\label{TPS1}
\partial_{\rho}f^{(1)}_{E} &=& -{1\over 2} {(g^{E(1)}_{00})^2 g^{E(1)}_{YY}\over \rho} \partial_{z}{\cal A}^{(1)}_{E} ,\\
\partial_{z} f^{(1)}_{E} &=& {1\over 2} {(g^{E(1)}_{00})^2 g^{E(1)}_{YY}\over \rho} \partial_{\rho}{\cal A}^{(1)}_{E},
\end{eqnarray}

{\it $\Omega^{(i)}_{E}$ satisfy }

\begin{eqnarray}\label{OS1}
\rho^{-1}\partial_{\rho}\Omega^{(i)}_{E} &=& {3\over 16}\left[\left(\partial_{\rho} \ln\left( g^{E(i)}_{YY}\right)\right)^2
- \left(\partial_{z} \ln \left(g^{E(i)}_{YY}\right)\right)^2 \right],\\
\rho^{-1}\partial_{z}\Omega^{(i)}_{E} &=&
{3\over 8} \partial_{\rho} \ln \left(g^{E(i)}_{YY}\right)\partial_{z}\ln\left( g^{E(i)}_{YY}\right),
\end{eqnarray}

{\it and  $\nu_{\zeta }$  solves the equations}

\begin{eqnarray}
\rho^{-1}\partial_{\rho}\nu_{\zeta} &=& {1\over 4}\left[\left(\partial_{\rho} \zeta\right)^2
- \left(\partial_{z} \zeta \right)^2 \right],\\
\rho^{-1}\partial_{z} \nu_{\zeta}&=& {1\over 2} \partial_{\rho} \zeta\partial_{z} \zeta.
\end{eqnarray}

Let us also note that, in general, the exchange of the two sigma models $M^{(1)} \longleftrightarrow M^{(2)}$
leads to different EMd solutions.

The presented proposition gives us a tool to generate new 5D EMd solutions in a
simple way  from
known solutions to the vacuum 5D Einstein equations. The technical difficulties are eventually
concentrating in finding of $\Omega_{E}$, $f_{E}$ and $\nu_{\zeta}$, but there are no principle obstacles
since the systems for $\Omega_{E}$, $f_{E}$ and $\nu_{\zeta}$ are first order partial differential equation systems of
the simplest kind.

Through the use of the proposition we can generate the "5D EMd images" of all known solutions of the vacuum
5D Einstein equations with the symmetries we consider here.
It is not possible to  present explicitly here  the "EMd images" of all
known vacuum Einstein solutions. We shall consider here one of the  most interesting
examples--namely we shall derive the rotating dipole black ring solutions.

\section{ Derivation of the rotating dipole black ring  \\ solutions}

The systematic derivation of the dipole black ring solutions in EM gravity was given in our previous paper
\cite{Y}. More precisely, we have shown that the EM dipole rings can be derived as a "nonlinear superposition"
of two neutral rotating black rings solutions. Here, we shall follow the same scheme
in order to derive the EMd rotating dipole ring solutions.

We take two copies of the neutral black ring solution with different parameters:
the first solution is with parameters $\{\lambda_{1},\nu,{\cal R}\}$ while the second is parameterized
by $\{\lambda_{2},\nu,{\cal R}\}$:

\begin{eqnarray}
ds^2_{E(i)} = -{F_{\lambda_{i}}(y)\over F_{\lambda_{i}}(x)} \left(dt + C(\nu,\lambda_{i})
{\cal R}{1+y\over F_{\lambda_{i}}(y)}d\psi \right)^2 \nonumber\\
+ {{\cal R}^2\over (x-y)^2 }F_{\lambda_{i}}(x)\left[-{G(y)\over F_{\lambda_{i}}(y)}d\psi^2 - {dy^2\over G(y)} + {dx^2\over G(x)}
+ {G(x)\over F_{\lambda_{i}}(x)}d\phi^2 \right]
\end{eqnarray}

where

\begin{equation}
F_{\lambda_{i}}(x) = 1 + \lambda_{i} x ,\,\,\, G(x) = (1-x^2)(1+\nu x),
\end{equation}

and

\begin{equation}
C(\nu,\lambda_{i})= \sqrt{\lambda_{i}(\lambda_{i}-\nu){1 +\lambda_{i} \over 1-\lambda_{i}}}.
\end{equation}

It should be also noted that in the case under consideration
the Killing vectors are denoted by

\begin{equation}
K_{1} = {\partial/\partial \psi} , \,\,\,\, K_{2} = {\partial/\partial \phi}.
\end{equation}

The neutral black ring solution has already been written  in canonical coordinates in \cite{HAR},
that is why we present here the final formulas:

\begin{eqnarray}
|g^{E(i)}_{00}| &=& {(1+\lambda_{i})(1-\nu)R_{1} + (1-\lambda_{i})(1+\nu)R_{2}
-2(\lambda_{i} - \nu)R_{3}
- \lambda_{i}(1-\nu^2){\cal R}^2  \over (1+\lambda_{i})(1-\nu)R_{1}
+ (1-\lambda_{i})(1+\nu)R_{2}
-2(\lambda_{i} - \nu)R_{3}
+ \lambda_{i}(1-\nu^2){\cal R}^2 } , \nonumber \\
g^{E(i)}_{\phi\phi} &=& {(R_{3}+z - {1\over 2}{\cal R}^2 )(R_{2} - z
+ {1\over 2}{\cal R}^2\nu)
\over R_{1} - z - {1\over 2}{\cal R}^2\nu }  \nonumber \\ &=& {(R_{1} + R_{2} + \nu{\cal R}^2)  (R_{1} - R_{3}
+ {1\over 2}(1+ \nu){\cal R}^2) (R_{2} + R_{3} - {1\over 2}(1 - \nu){\cal R}^2)
\over {\cal R}^2 ((1-\nu)R_{1} - (1+\nu)R_{2} -2\nu R_{3}) } ,\nonumber \\
 g^{E(i)}_{\rho\rho} &=&  [(1+\lambda_{i})(1-\nu)R_{1} + (1-\lambda_{i})(1+\nu)R_{2}
 -2(\lambda_{i} - \nu)R_{3}
+ \lambda_{i}(1-\nu^2){\cal R}^2 ] \nonumber \\
&& \times {(1-\nu)R_{1} + (1+\nu)R_{2} + 2\nu R_{3}
\over  8(1-\nu^2)^2 R_{1}R_{2}R_{3}} , \\
{\cal A}^{(i)}_{E} &=& {-2 C(\nu,\lambda_{i}) {\cal R} (1-\nu)
[R_{3} -R_{1} + {1\over 2}{\cal R}^2 (1+\nu)] \over
(1+\lambda_{i})(1-\nu)R_{1} + (1-\lambda_{i})(1+\nu)R_{2} -2(\lambda_{i} - \nu)R_{3}
- \lambda_{i}(1-\nu^2){\cal R}^2 },  \nonumber
\end{eqnarray}

where

\begin{eqnarray}
R_{1} =\sqrt{\rho^2 + (z + {\nu\over 2}{\cal R}^2)^2 } , \\
R_{2} =\sqrt{\rho^2 + (z - {\nu\over 2}{\cal R}^2)^2 }, \\
R_{3} = \sqrt{\rho^2 + (z - {1\over 2}{\cal R}^2)^2 }.
\end{eqnarray}

The next step is to find the functions $\Omega^{(i)}_{E}$ and $f^{(i)}_{E}$. After straightforward but tedious
calculations we obtain

\begin{eqnarray}
e^{{8\over 3} \Omega_{E}^{(1)}} &=& e^{{8\over 3} \Omega_{E}^{(2)}} = { [(1-\nu)R_{1} + (1+\nu)R_{2}
+ 2\nu R_{3}]^2\over 8(1-\nu^2)^2R_{1}R_{2}R_{3} }
g^{E(i)}_{\Phi\Phi} ,\\
f^{(i)}_{E} &=& {(1-\nu) {\cal R} C(\nu,\lambda_{i}) [R_{1} - R_{3} +
{1\over 2}(1+\nu ) {\cal R}^2 ] \over (1+\lambda_{i})(1-\nu)R_{1} +
(1-\lambda_{i})(1+\nu)R_{2} + 2(\nu-\lambda_{i})R_{3}
+ \lambda_{i}(1-\nu^2){\cal R}^2 } \nonumber .
\end{eqnarray}

Finally we have to choose a harmonic function $\zeta$. It turns out that the appropriate
choice is

\begin{equation}
\zeta  = u^{(1)}_{E} =u^{(2)}_{E} = {1\over 2}\ln(g^{E(1)}_{\phi\phi})={1\over 2}\ln(g^{E(2)}_{\phi\phi}).
\end{equation}

With this choice we find\footnote{We have taken into account that $g^{E(1)}_{\phi\phi}=g^{E(2)}_{\phi\phi} $
and $\Omega^{(1)}_{E}= \Omega^{(2)}_{E}=3\nu_{\zeta}$ which considerably simplifies the solution.  }

\begin{eqnarray}
ds^2 &=& |g^{E(1)}_{00}|^{2\over 1+ \alpha^2_{*}} g^{E(1)}_{\phi\phi} d\phi^2 +
|g^{E(1)}_{00}|^{-{1\over 1+ \alpha^2_{*}}}
\left[g^{E(2)}_{00}\left(dt + {\cal A}^{E(2)}d\psi \right)^2  + {\tilde g}^{E(2)}_{\psi\psi} d\psi^2
 \right. \nonumber \\
&& \left.
+    \left(|g^{E(1)}_{00}|g^{E(1)}_{\phi\phi}g^{E(1)}_{\rho\rho} \over e^{{8\over 3}\Omega^{(1)}_{E} } \right)^{3\over 1+\alpha^2_{*}}
 g^{E(2)}_{\rho\rho} (d\rho^2 + dz^2) \right] ,\\
A_{\phi} &=& \pm {2\sqrt{3}\over \sqrt{1+ \alpha^2_{*}}} f^{(1)}_{E} + const, \\
e^{2\alpha\varphi} &=& |g^{E(1)}_{00}|^{2\alpha^2_{*}\over 1+ \alpha^2_{*}} .
\end{eqnarray}

It is more convenient to
present the solution in coordinates in which it takes  simpler form. Such coordinates are the so-called
$C$-metric coordinates given by

\begin{eqnarray}
\rho = {{\cal R}^2 \sqrt{-G(x)G(y)}\over (x-y)^2 } ,\,\,\,
z = {1\over 2} {{\cal R}^2(1-xy)(2+\nu x + \nu y )\over (x-y)^2 }.
\end{eqnarray}

Performing this coordinate change we find

\begin{eqnarray}
ds^2 &=& \left({F_{\lambda_{1}}(y)\over F_{\lambda_{1}}(x)}\right)^{2\over1+ \alpha^2_{*} } {{\cal R}^2G(x)\over (x-y)^2} d\phi^2
+  \left({F_{\lambda_{1}}(y)\over F_{\lambda_{1}}(x)}\right)^{-{1\over1+ \alpha^2_{*}} }
\left[-{F_{\lambda_{2}}(y)\over F_{\lambda_{2}}(x)} \left(dt + C(\nu,\lambda_{2})
{\cal R}{1+y\over F_{\lambda_{2}}(y)}d\psi \right)^2 \right. \nonumber \\ && \left.
 -  {{\cal R}^2F_{\lambda_{2}}(x)\over (x-y)^2 }{G(y)\over F_{\lambda_{2}}(y)}d\psi^2
+ \left(F_{\lambda_{1}}(y) \right)^{3\over 1+\alpha^2_{*}} {{\cal R}^2F_{\lambda_{2}}(x)\over (x-y)^2 }
\left(- {dy^2\over G(y)} + {dx^2\over G(x)}\right)  \right] ,\\
A_{\phi} &=& \pm {\sqrt{3}C(\nu,\lambda_{1})\over \sqrt{1+ \alpha^2_{*} }}{\cal R} {1+x\over F_{\lambda_{1}}(x) } + const ,\\
e^{2\alpha\varphi} &=&
\left({F_{\lambda_{1}}(y)\over F_{\lambda_{1}}(x)}\right)^{2\alpha^2_{*}\over1+ \alpha^2_{*} } .
\end{eqnarray}

Finally, in order to exclude pathological behavior of the metric and to obtain black solutions we must consider only negative $\lambda_{1}$, i.e.

\begin{equation}
\lambda_{1} = - \mu \,\,\, , 0\le\mu<1 .
\end{equation}

and positive $\lambda_{2}$ and $\nu$ satisfying

\begin{equation}
0<\nu \le \lambda_{2} <1.
\end{equation}

One can easily see that the generated 5D EMd solutions are just the EMd rotating dipole black ring solutions \cite{EMP} .
Let us also recall\cite{EMP} that in order to avoid the possible conical singularities at $x=\pm 1$ and $y=-1$
we must impose

\begin{eqnarray}
&&\Delta \phi = \Delta \psi = 2\pi {(1 + \mu)^{3\over 2(1+ \alpha^2_{*})} \sqrt{1-\lambda_{2}}\over 1-\nu } ,\\
&&{1-\lambda_{2}\over 1+\lambda_{2} } \left( {1+ \mu\over 1-\mu } \right)^{3\over 1+ \alpha^2_{*}} = \left({1-\nu\over 1+\nu } \right)^2 .
\end{eqnarray}

An alternative derivation of the the rotating dipole black ring solutions is given in the Appendix.

\section{Conclusion}

In this paper we considered EMd gravity in spacetimes admitting three commuting Killing vectors: one timelike and
two spacelike one of them being hypersurface orthogonal. Assuming also a special ansatz for the electromagnetic
field we have shown that  the EMd equations reduce to one linear equation, two $SL(2,R)/SO(2)$ $\sigma$-models  and
a separated linear system of first order partial differential equations. This ensures the existence of Lax-pair
presentation, therefore the complete integrability of the considered sector of EMd gravity.
The Lax pair presentation also opens the way to apply the IST method and  to generate multisoliton
solutions.

Using the two $\sigma$-models structure of the reduced EMd sector we gave an explicit construction
for generating exact 5D EMd solutions from known solutions of the 5D vacuum Einstein equations in the same
symmetry sector. As an example we gave, for the first time, the explicit and systematic
derivation of the rotating dipole black ring solutions.

The presented solution generating method can also be used to generate many other exact 5D EMd solutions.
It would be interesting to find the EMd solutions corresponding to the "nonlinear superposition"
of 5D Myers-Perry black holes \cite{MPER} as well as other 5D vacuum Einstein solutions.

\section*{Acknowledgements}
I would like to thank I. Stefanov for reading the manuscript.
This work was partially supported by the
Bulgarian National Science Fund under Grant MUF04/05 (MU 408)
and the Sofia University Research Fund under Grant No60.

\section*{Appendix}

In many cases it turns out that the canonical coordinates are not the most
convenient ones and the corresponding equations are more tractable in other coordinates.
Here we present the basic results in terms of the coordinates $x$ and $y$ in which the transverse space metric
has the  form

\begin{equation}
dl^2 = e^{-2U(x,y)}e^{2H(x,y)}\left({dx^2\over A(x) } + {dy^2\over B(y)} \right)
\end{equation}

where $A(x)$ and $B(y)$ are appropriate functions.

For the reduced EMd equations we find

\begin{eqnarray}
&&\partial_{x}\left[\sqrt{A(x)\over B(y)}\partial_{x}\rho\right] +
\partial_{y}\left[\sqrt{B(y)\over A(x)}\partial_{y}\rho\right]=0 ,
\end{eqnarray}
\begin{eqnarray}
&&\partial_{x}\left[\rho\sqrt{A(x)\over B(y)}\partial_{x}M_{1}M^{-1}_{1}\right] +
\partial_{y}\left[\rho\sqrt{B(y)\over A(x)}\partial_{y}M_{1}M^{-1}_{1}\right]=0,
\end{eqnarray}

\begin{eqnarray}
&&\partial_{x}\left[\rho\sqrt{A(x)\over B(y)}\partial_{x}M_{2}M^{-1}_{2}\right] +
\partial_{y}\left[\rho\sqrt{B(y)\over A(x)}\partial_{y}M_{2}M^{-1}_{2}\right]=0,
\end{eqnarray}

\begin{eqnarray}
&&\partial_{x}\left[\rho\sqrt{A(x)\over B(y)}\partial_{x}\zeta\right] +
\partial_{y}\left[\rho\sqrt{B(y)\over A(x)}\partial_{y}\zeta\right]=0,
\end{eqnarray}

\begin{eqnarray}
&&{S\over \rho}\partial_{x}H = {1\over 2\rho}\partial_{x}S +
{1\over 8}\left[B(y)\partial_{x}\rho Tr\left(\partial_{y} M_{2}\partial_{y} M^{-1}_{2} \right)
- A(x)\partial_{x}\rho Tr\left(\partial_{x} M_{2}\partial_{x} M^{-1}_{2} \right) \right] \nonumber \\
&&+ {3\over 8(1+ \alpha^2_{*})}\left[B(y)\partial_{x}\rho Tr\left(\partial_{y} M_{1}\partial_{y} M^{-1}_{1} \right)
- A(x)\partial_{x}\rho Tr\left(\partial_{x} M_{1}\partial_{x} M^{-1}_{1} \right) \right] \nonumber \\
&& - {1\over 4} B(y)\partial_{y}\rho Tr\left(\partial_{x} M_{2}\partial_{y} M^{-1}_{2} \right)
- {3\over 4(1+ \alpha^2_{*})} B(y)\partial_{y}\rho Tr\left(\partial_{x} M_{1}\partial_{y} M^{-1}_{1} \right) \\
&&+ {3\alpha^2_{*}\over 4(1+\alpha^2_{*})}  \left(A(x)\partial_{x}\rho (\partial_{x}\zeta)^2  - B(y)\partial_{x}\rho
(\partial_{y}\zeta)^2 \right) + {3\alpha^2_{*}\over 2(1+\alpha^2_{*})} B(y)\partial_{y}\rho \partial_{x}\zeta \partial_{y}\zeta ,
\nonumber
\end{eqnarray}

\begin{eqnarray}
&&{S\over \rho}\partial_{y}H = {1\over 2\rho}\partial_{y}S +
{1\over 8}\left[A(x)\partial_{y}\rho Tr\left(\partial_{x} M_{2}\partial_{x} M^{-1}_{2} \right)
- B(y)\partial_{y}\rho Tr\left(\partial_{y} M_{2}\partial_{y} M^{-1}_{2} \right) \right] \nonumber \\
&&+ {3\over 8(1+ \alpha^2_{*})}\left[A(x)\partial_{y}\rho Tr\left(\partial_{x} M_{1}\partial_{x} M^{-1}_{1} \right)
- B(y)\partial_{y}\rho Tr\left(\partial_{y} M_{1}\partial_{y} M^{-1}_{1} \right) \right] \nonumber \\
&& - {1\over 4} A(x)\partial_{x}\rho Tr\left(\partial_{x} M_{2}\partial_{y} M^{-1}_{2} \right)
- {3\over 4(1+ \alpha^2_{*})} A(x)\partial_{x}\rho Tr\left(\partial_{x} M_{1}\partial_{y} M^{-1}_{1} \right) \\
&&+ {3\alpha^2_{*}\over 4(1+\alpha^2_{*})}  \left(B(y)\partial_{y}\rho (\partial_{y}\zeta)^2  - A(x)\partial_{y}\rho
(\partial_{x}\zeta)^2 \right) + {3\alpha^2_{*}\over 2(1+\alpha^2_{*})} A(x)\partial_{x}\rho \partial_{x}\zeta \partial_{y}\zeta
\nonumber ,
\end{eqnarray}

where

\begin{equation}
S = A(x)(\partial_{x}\rho)^2 + B(y)(\partial_{y}\rho)^2.
\end{equation}

The equations for the twist potential are

\begin{eqnarray}
\partial_{x}f &=& - {1\over 2}\sqrt{B(y)\over A(x)} {e^{4U}\over \rho } \partial_{y}{\cal A} ,\\
\partial_{y}f &=&  {1\over 2}\sqrt{A(x)\over B(y)} {e^{4U}\over \rho } \partial_{x}{\cal A} .
\end{eqnarray}

Further we proceed as in section 3. Potentials $u$, $A_{Y}$ and $\varphi$ are again given by (\ref{UAD}).
One can show that

\begin{eqnarray}\label{HD}
H= H^{(2)}_{E} - \Omega^{(2)}_{E} + {3\over 1+ \alpha^2_{*}} \left[H^{(1)}_{E} - {1\over 2}\ln(S) - \Omega^{(1)}_{E}  \right]
+  {3\alpha^2_{*}\over 1+ \alpha^2_{*}}\nu_{\zeta}
\end{eqnarray}

where $\Omega^{(i)}_{E}$ are solutions to the systems

\begin{eqnarray}\label{OXYE}
&&{S\over \rho}\partial_{x}\Omega^{(i)}_{E} = {3\over 4}\left[A(x)\partial_{x}\rho (\partial_{x}u^{(i)}_{E})^2 -
B(y)\partial_{x}\rho (\partial_{y}u^{(i)}_{E})^2 \right]  +
{3\over 2}B(y)\partial_{y}\rho \partial_{x}u^{(i)}_{E}\partial_{y}u^{(i)}_{E} ,\\
&&{S\over \rho}\partial_{y}\Omega^{(i)}_{E} = {3\over 4}\left[B(y)\partial_{y}\rho (\partial_{y}u^{(i)}_{E})^2 -
A(x)\partial_{y}\rho (\partial_{x}u^{(i)}_{E})^2 \right]  +
{3\over 2}A(x)\partial_{x}\rho \partial_{x}u^{(i)}_{E}\partial_{y}u^{(i)}_{E} , \nonumber
\end{eqnarray}

and $\nu_{\zeta}$ satisfies

\begin{eqnarray}
&&{S\over \rho}\partial_{x}\nu_{\zeta} = {1\over 4}\left[A(x)\partial_{x}\rho (\partial_{x}\zeta)^2 -
B(y)\partial_{x}\rho (\partial_{y}\zeta)^2 \right]  +
{1\over 2}B(y)\partial_{y}\rho \partial_{x}\zeta\partial_{y}\zeta ,\\
&&{S\over \rho}\partial_{y}\nu_{\zeta}= {1\over 4}\left[B(y)\partial_{y}\rho (\partial_{y}\zeta)^2 -
A(x)\partial_{y}\rho (\partial_{x}\zeta)^2 \right]  +
{1\over 2}A(x)\partial_{x}\rho \partial_{x}\zeta\partial_{y}\zeta . \nonumber
\end{eqnarray}

Taking into account the explicit expression of $e^{2u}$ from (\ref{UAD})  and the decomposition (\ref{HD})
we find for the EMd metric the following formula

\begin{eqnarray}
ds^2 = e^{4U^{(1)}_{E}\over 1 + \alpha^2_{*}} e^{{2\alpha^2_{*}\over 1 + \alpha^2_{*}}\zeta } dY^2 +
e^{-{2U^{(1)}_{E}\over 1 + \alpha^2_{*}}} e^{-{\alpha^2_{*}\over 1 + \alpha^2_{*}}\zeta }
\left[ -e^{2U^{(2)}_{E}} \left(dt + {\cal A}^{(2)}_{E} dX \right)^{2}  + e^{-2U^{(2)}_{E}}\rho^2 dX^2
\right. \nonumber \\
\left. + \left(S^{-1} e^{2H^{(1)}_{E}} \over e^{2\Omega^{(1)}_{E}+ {2\over 3}\Omega^{(2)}_{E}} \right)^{3\over 1 + \alpha^2_{*} }
\left( {e^{2\nu_{\zeta}}  \over e^{{2\over 3}}\Omega^{(2)}_{E} }  \right)^{3\alpha^2_{*}\over 1 + \alpha^2_{*} }
e^{-2U^{(2)}_{E}}e^{2H^{(2)}_{E} }\left({dx^2\over A(x) } + {dy^2\over B(y) } \right)  \right] .
\end{eqnarray}

The above presented results  give us an opportunity to derive dipole black solutions more simply when the
the transverse space coordinates are appropriately chosen. The most natural and convenient choice is

\begin{eqnarray}
A(x) &=& G(x),\\
B(x) &=& - G(y).
\end{eqnarray}

Then, from the neutral black ring solution we find

\begin{eqnarray}\label{EFUUARH}
&&e^{2u^{(i)}_{E}}={{\cal R}^2 G(x)\over (x-y)^2 },\nonumber\\
&&{\cal A}^{(i)}_{E} = C(\nu,\lambda_{i}){\cal R}{1+y\over F_{\lambda_{i}}(y)}, \nonumber \\
&&e^{4U^{(i)}_{E}} = {F^2_{\lambda_{i}}(y)\over F^2_{\lambda_{i}}(x) } {{\cal R}^2 G(x)\over (x-y)^2} ,\\
&&\rho = {{\cal R}^2 \over (x-y)^2} \sqrt{-G(x)G(y)}, \nonumber \\
&& e^{2H^{(i)}_{E}} = {{\cal R}^4 F_{\lambda_{i}}(y) G(x) \over (x-y)^4 } \nonumber .
\end{eqnarray}

Having $A(x)$, $B(y)$ and $\rho$ we can calculate $S$ and the result is

\begin{eqnarray}
S =  {\cal R}^4 { (x+y +\nu +\nu xy) \over 4(x-y)^3}\left[2-\nu +\nu(x+y) + \nu xy  \right]
\left[-2-\nu - \nu(x+y) + \nu xy  \right].
\end{eqnarray}

Solving  equations (\ref{OXYE}) we find

\begin{equation}
e^{{8\over 3}\Omega^{(1)}_{E}(x,y)} = e^{{8\over 3}\Omega^{(2)}_{E}(x,y)} =  S^{-1}(x,y) {{\cal R}^4 G(x)\over (x-y)^4 }.
\end{equation}

Therefore we have

\begin{eqnarray}
{S^{-1}e^{2H^{(1)}_{E}} \over e^{{8\over 3}\Omega^{(1)}_{E}}  } = F_{\lambda_{1}}(y) .
\end{eqnarray}

The next step is to specify the harmonic function $\zeta$. As we have already mentioned in section 4 the appropriate choice is

\begin{equation}
\zeta = u^{(1)}_{E} = u^{(2)}_{E}
\end{equation}

which means that $\nu_{\zeta}= {1\over 3}\Omega^{(1)}_{E}$.

Summarizing, we obtain the following expression for the 5D EMd metric

\begin{eqnarray}
ds^2 &=& \left( {F_{\lambda_{1}}(y)\over F_{\lambda_{1}}(x) }\right)^{2\over 1 + \alpha^2_{*}} {{\cal R}^2G(x)\over (x-y)^2 }d\phi^2 + \left( {F_{\lambda_{1}}(y)\over F_{\lambda_{1}}(x) }\right)^{-{1\over 1 + \alpha^2_{*}}}
\left[- {F_{\lambda_{2}}(y)\over F_{\lambda_{2}}(x) }\left(dt + C(\nu,\lambda_{2}){\cal R} {1+y\over F_{\lambda_{2}}(y)}d\psi \right)^2  \right. \nonumber \\
&& \left.  - {{\cal R}^2 G(y)\over (x-y)^2 } {F_{\lambda_{2}}(x)\over  F_{\lambda_{2}}(y) }d\psi^2  +
\left(F_{\lambda_{1}}(y)  \right)^{3\over 1+ \alpha^2_{*}} {{\cal R}^2 F_{\lambda_{2}}(x) \over (x-y)^2} \left({dx^2\over G(x) }
- {dy^2\over G(y)} \right) \right] .
\end{eqnarray}

Taking into account the explicit form of $e^{4U^{(1)}_{E}}$, ${\cal A}^{(1)}_{E}$ and $\rho$ from (\ref{EFUUARH}),
the equations for the twist potential become

\begin{eqnarray}
\partial_{x}f^{(1)}_{E} &=& -{1 \over 2} C(\nu,\lambda_{1}) {{\cal R} (1-\lambda_{1})\over F^{2}_{\lambda_{1}}(x) } ,\\
\partial_{y}f^{(1)}_{E} &=& 0.
\end{eqnarray}

Integrating we obtain

\begin{equation}
f^{(1)}_{E} = -{1 \over 2} C(\nu,\lambda_{1}) {{\cal R} (1 + x)\over F_{\lambda_{1}}(x)}  + const.
\end{equation}

Therefore the electromagnetic field is given by

\begin{eqnarray}
A_{Y} = \pm {\sqrt{3} C(\nu,\lambda_{1})  \over \sqrt{1 + \alpha^2_{*}}} {\cal R} {(1 + x)\over F_{\lambda_{1}}(x)}  + const.
\end{eqnarray}

\end{document}